\documentclass{ismdproc}

\begin{document}

\title{Theory of transverse-momentum parton densities: solving the puzzle of divergences}
\author{{\slshape I.O. Cherednikov\footnote{Also at: {\sl ITPM,
             Moscow State University, Russia}; Email: {\tt igor.cherednikov@ua.ac.be} }}  \\[1ex]
Universiteit Antwerpen, Groenenborgerlaan 171, 2020 Antwerpen, Belgium \\
Bogoliubov Laboratory of Theoretical Physics, JINR, 141980 Dubna, Russia}

\contribID{xy}  
\confID{yz}
\acronym{ISMD2010}
\doi            

\maketitle

\begin{abstract}
The current status of the theoretical understanding of the transverse-momentum dependent parton densities (TMDs) is discussed.
Special attention is payed to the difference between the operator definitions of TMDs proposed so far, the treatment of
specific divergences, the geometry of the gauge links, and the role of the soft factors.
\end{abstract}

Short-distance factorization is the basic concept for the application of QCD at high energies \cite{ER80Riv, CSS89, CTEQ}.
The predictive power of QCD in this region has been successfully demonstrated in the case of fully inclusive processes, where the collinear (Feynman) parton distribution functions (PDFs) are applicable as a nonperturbative part of the QCD factorization theorems.
Application of the QCD factorization approach to the description of the semi-inclusive hadronic reactions, such as the semi-inclusive DIS, or the Drell-Yan (DY) process, demands an expansion of the concept of parton distribution functions beyond the collinear approximation. It is useful to introduce the transverse-momentum-dependent (TMD) parton distribution (or fragmentation) functions (in what follows I abbreviate them by TMD = transverse-momentum ``densities'') are essentially non-perturbative objects, which accumulate information about the intrinsic 3-dimensional motion of partons (quarks and/or gluons) in a hadron.
In Refs. \cite{Sop77, Col78, RS79}, the TMDs have been proposed as a generalization of the collinear PDFs:
\begin{equation}
 {F}_{\rm [collinear]} (x, \mu)  \ \to \ {\cal F}_{\rm [tmd]} (x, \mathbf k_\perp, \mu, \zeta) \ .
 \label{eq:tmd_soper}
\end{equation}
It was pointed out that, in contrast to the collinear case, TMD depends on ``how fast the hadron in moving'', that is, on  hadron's rapidity. Formally, this dependence enters the Eq. (\ref{eq:tmd_soper}) via the additional variable $\zeta =
(2 P\cdot v)/ |v^2| \ , \ v^2 \neq 0$, where $P$ is the incoming nucleon momentum and $n$ is non-light-like axial gauge fixing vector. The appearance of the additional rapidity dependence in the TMDs requires the development of an appropriate resummation scheme and the derivation of suitable evolution equations.
Another important consequence of the rapidity dependence that has been made by Soper \cite{Sop77}: the very definition of a {\it parton} is what changes as rapidity varies. In the collinear case, the only extra variable is the renormalization scale $\mu$, and the gauge-invariant operator definition of integrated PDFs allows one to relate their moments to the matrix elements of the twist-two operators arising in the operator product
expansion on the light-cone \cite{CS81}, thus making the PDFs consistent from the point of view of the local quantum field theory and allowing the probabilistic interpretation in terms of the parton number operators. The renormalization properties of the collinear PDFs are governed by the DGLAP equation, establishing the logarithmic dependence of the structure functions on the hard scale $Q^2$.

Need of a consistent theory of the TMDs as quantum objects initiated the quest for the deeper understanding of such fundamental issues as QCD factorization, universality of PDFs, their renormalizability and complete gauge invariance. The next important step in the theory of TMDs had been done in the works \cite{CS81, CS82}: it was observed that trying to write down the operator definition of the (quark) TMD in the form similar to the collinear PDF (the axial gauge $(A\cdot v) = 0\ , \ v^2 \neq 0 $ is used to remove the longitudinal gauge links)
\begin{equation}
  {\cal F}_{\rm i/h} (x, \mathbf k_\perp, \mu, \zeta)
=
  \frac{1}{2}
  \int \frac{d\xi^- d^2\mbox{\boldmath$\xi_\perp$}}{2\pi (2\pi)^2}
  {\rm e}^{-ik^{+}\xi^{-} +i \mbox{\footnotesize\boldmath$k_\perp$}
\cdot \mbox{\footnotesize\boldmath$\xi_\perp$}}
  \left\langle
              h |\bar \psi_i (\xi^-, \mbox{\boldmath$\xi_\perp$})
              \gamma^+
   \psi_i (0^-,\mbox{\boldmath$0_\perp$}) | h
   \right\rangle \ \, ,
\label{eq:tmd_naive}
\end{equation}
where the extra parameter is defined via the gauge-fixing vector $\zeta = (2 P \cdot v)^2/|v^2|$, one
is able to separate out the emergent rapidity divergences in the form of the (powers of) logarithms $\ln \zeta$ and
perform their resummation by means of suitable evolution equation \cite{CS81}. Although the formal integration of
definition (\ref{eq:tmd_naive}) over $\mathbf k_\perp$ yields the collinear PDF
\begin{equation}
  \int\! d^2 k_\perp \ {\cal F}_{\rm i/h} (x, \mathbf k_\perp, \mu, \zeta)
  \to
 { {F}_{\rm [collinear]} (x, \mu) } \ ,
\end{equation}
such a transition can only be justified in the tree approximation, since the presence of the rapidity divergences
breaks down this direct procedure. Nevertheless, important progress has been achieved in the semi-inclusive factorization (in particular, for the Drell-Yan process) in the works \cite{CS81, CS82} and \cite{CSS85}.

Since the beginning of the nineties, a wide range of the phenomenological application of the TMDs has been revealed in the polarized hadronic processes. In particular, it was found that TMDs play a crucial role in the understanding of the single-spin asymmetries (see, e.g., Ref. \cite{Siv90, Kotz94, TM_all, Torino, Bacch07, polarized_rev} and Refs. therein). It has been realized that a much deeper knowledge of the properties of the polarized and unpolarized TMDs is essential not only because of the great theoretical interests in them, but from pragmatic needs: one has to be able to use the correct operator definitions of TMDs, to know the complete set of their evolution equations, to have justified factorization theorems, to keep under control the universality of the TMDs, etc. Since then, important steps have been done in the study of the initial and final state interactions in the operator formulation of TMDs and their relations with the structure of Wilson lines \cite{JY02, BJY03, BMP03}; TMD factorization in the covariant gauges \cite{JMY04}; treatment of the extra singularities within the subtraction scheme \cite{CH00, Hau07}; problems of the universality and gauge invariance of TMDs and structure of the gauge links \cite{CM04, BoM07}; the problem of (breakdown of) factorization in certain semi-inclusive processes \cite{CQ07, CRS07, RM10};  problems related to the emergent self-energy singularities of the Wilson lines entering the definition of TMDs \cite{Col08}; the issue of matching different methods of taking into account the $\mathbf k_\perp$-dependence at different momentum scales \cite{Bacch08}; various aspects of the evolution equations for the TMDs \cite{CT_ev, AR11}; calculations of the TMD within the quark models \cite{Q_Models}, development of the lattice simulations methods for the TMDs \cite{Lat_TMD}, development a generalized approach to the operator definition of the TMDs taking into account their renormalization-group properties in the light-cone gauge \cite{CS_all, SC09, CKS10}, etc. Further important references can be found, e.g., in Refs. \cite{Col03, BR05}.

The main problem I intend to discuss in the present work is the treatment of pathological singularities beyond the tree-approximation in the TMDs.
To be precise, at the one-loop level the following sorts of singular terms arise:
\begin{itemize}

  \item {\it Ultraviolet poles} ${\sim \frac{1}{\varepsilon}}$ in the
  dimensional regularization: they have to be removed by the standard renormalization procedure.

  \item {\it Rapidity divergences}:
  they depend on the additional rapidity parameter $\zeta$ \cite{CS81, CS82}, but do not break the renormalizability of the TMDs, and can be safely resummed by means of the Collins-Soper equation.

  \item Pathological {\it overlapping divergences}: they contain the UV
  and rapidity poles simultaneously
  $
  {\sim \frac{1}{\varepsilon} \ \ln \eta \ }\, .
  $
  They  break down standard renormalizability of TMDs, calling for
  a {generalized} renormalization procedure
  in order to enable the construction of consistent operator definition of the TMDs.

\end{itemize}

To overpass the problem of emergent extra divergences, different approaches to the operator definition of TMDs have been
proposed. In what follows, my notations are: A or C denote the {\it axial} or {\it covariant} gauge, respectively, and the subscripts $v$ or $n$ mark the vectors defining the {\it non-light-like} or {\it light-like} directions of the longitudinal gauge links (in covariant gauges) or the gauge-fixing vector (in axial gauges), respectively. I will distinguish between the following definitions of TMDs:

\begin{itemize}

  \item A$_{\rm v}$-TMD: one applies the axial non-light-like gauge, the longitudinal (along the vector $v$) gauge links vanish by virtue that $(v \cdot A) = 0$; the rapidity cutoff is defined as $\zeta = (2 P\cdot v)^2/|v^2|$, Refs. \cite{CS82, CSS85}.

  \item C$_{\rm v}$-TMD: in the covariant gauge, the longitudinal gauge links along the non-light-like vector survive, the transverse gauge links cancel; the rapidity cutoff $\zeta = (2 P\cdot v)^2/|v^2|$ characterizes the deviation of the longitudinal gauge links from the light-like direction; the soft factor contains the non-light-like gauge links, Refs. \cite{JMY04}.

  \item A$_{\rm n}$-TMD: one uses the light-like axial gauge $(n \cdot A) = 0 \ , \ n^2 = 0$, the light-like longitudinal gauge links vanish, the transverse gauge links at the light-cone infinity survive; the regularization parameter is defined as $\eta_{\rm LC} = (P \cdot n)/\eta$, where $\eta$ originates from
      the regularization of the $q^+$-pole in the gluon propagator, e.g., with the PV prescription
      \begin{equation}
      \frac{1}{[q^+]_{\eta} }
       =  \frac{1}{2}\left( \frac{1}{q^+ + i \eta}
                             +\frac{1}{q^+ - i \eta} \right) \ ,
                             \label{eq:prop_reg}
      \end{equation}
      the soft factor contains the light-like longitudinal and the transverse gauge links, Refs. \cite{CS_all}.

  \item C$_{\rm n}$-TMD: covariant gauge with light-like gauge links, no transverse gauge links; soft factor contains both light-like and non-light-like longitudinal, as well as the transverse gauge links; regularization parameter $\zeta_{\rm } = (2 P\cdot v)^2/|v^2|$, where $v$ is the non-light-like direction of gauge links in the soft factor, Refs. \cite{CH00, Hau07, CM04, Col03}.

  \item L-TMD: this definition is proposed for the lattice simulations; the direct connector is used as a gauge link, no regularization parameters, no light-like gauge links \cite{Lat_TMD}.

\end{itemize}
It is worth noting that there is no {\it a priori} direct relationship between these definitions, they define, in principle, different objects.

Let us discuss the properties of the emergent singularities within the definitions listed above.
Within the ``off-light-cone'' frameworks A$_{\rm v}$-TMD and C$_{\rm v}$-TMD, a rapidity divergence arises in the form of the logarithmic terms $\ln \zeta$ and $\ln^2 \zeta$. In the ``light-cone'' definitions A$_{\rm n}$-TMD and C$_{\rm n}$-TMD, the situation differs from the previous case.  It was shown in \cite{CS_all} that in the A$_{\rm n}$-TMD, the
anomalous divergent term containing overlapping (${\rm UV} \otimes {\rm rapidity}$ singularity) stems from the
virtual-gluon contribution (we consider the ``distribution of a quark in a quark with momentum $p$'' and the TMD is normalized as ${\cal F}_{\rm i/p}^{[{\rm A}_{\rm n}] (0)} (x, {\mathbf k_\perp}) = \delta(1 - x ) \delta^{(2)} ({\mathbf k_\perp})$)
\begin{equation}
     {\Sigma} [{\rm A}_{\rm n}]
     =
     - \frac{\alpha_s}{\pi} C_{\rm F} \  \Gamma(\epsilon)\
  \left[ 4 \pi \frac{\mu^2}{-p^2} \right]^\epsilon\
  \delta (1-x) \delta^{(2)} ({\boldmath k_\perp})\ \int_0^1\!
  dx \frac{(1-x)^{1-\epsilon}}{x^\epsilon [x]_\eta} \ ,
\label{eq:sigma-lc}
\end{equation}
where the $x$-integral is regularized according to the rule (\ref{eq:prop_reg}).
In the covariant gauge, the definition C$_{\rm n}$-TMD yields \cite{CS82}
\begin{equation}
 \Sigma[{\rm C}_{\rm n}]
 =
 - \frac{\alpha_s}{\pi} C_{\rm F} \Gamma (\epsilon) \left[ 4\pi \frac{\mu^2}{-p^2} \right]^{\epsilon} \
 \delta (1-x) \delta^{(2)} ({\boldmath k_\perp} )\ \int_0^1\! dx \ \frac{x^{1-\epsilon}}{(1-x)^{1+\epsilon}} \ .
 \label{eq:sigma-cov}
\end{equation}
One easily observes that without
the $\eta$-regularization of the last integral and after a trivial change
of variables, Eq.\ (\ref{eq:sigma-lc}) is reduced to Eq.\ (\ref{eq:sigma-cov}):
$
  \Sigma[{\rm C}_{\rm n}] (\epsilon)
  =
  \Sigma[{\rm A}_{\rm n}] (\epsilon, \eta = 0) \ .
$
In principle, one can use dimensional regularization to take care of the overlapping singularities as well.
I will not discuss here this approach. In more detail, the applicability of dimensional regularization
for a consistent treatment of the divergences arising in the path-dependent
gauge invariant 2-quark correlation functions had been studied in Ref.\ \cite{Ste83}.

Within the ${\rm A}_{\rm n}$ and ${\rm C}_{\rm n}$, the cancellation of the overlapping divergences is achieved by means of the subtraction of the special {\it soft factors}. This procedure implies the generalized renormalization of the TMDs and provides us with a completely gauge invariant object, free of the overlapping divergences \cite{CS_all, SC09}.
However, another problem arises in the one-loop corrections to the soft factor itself. In the light-cone gauge, one obtains the term
\begin{equation}
  \Sigma_{\rm soft}[{\rm A}_{\rm n}]
=
  i g^2 \mu^{2\epsilon} C_{\rm F}2 p^+ \ \int\! \frac{d^\omega q}{(2\pi)^\omega}
  \frac{1}{q^2 (q^- \cdot p^+ - i0) [q^+]_\eta} \ ,
\label{eq:soft}
\end{equation}
where a new singularity emerges, that can not be removed neither by
dimensional regularization, nor by the $\eta$-cutoff. One has
\begin{equation}
  \Sigma_{\rm soft}[{\rm A}_{\rm n}]
=
  - \frac{\alpha_s}{\pi} C_{\rm F}  \left[\frac{4\pi \mu^2}{\lambda^2}\right]^\epsilon
  \Gamma(\epsilon) \
  \int_0^1 dx \frac{x}{x^2 [ x-1 ]_\eta} \ ,
\label{eq:soft_int}
\end{equation}
where $\lambda$ is the IR regulator.
Taking into account that the extra singularity is cusp (or rapidity) -independent \cite{CS_all},
we conclude that it represents the {\it self-energy of the Wilson line}
evaluated along a ``straightened'' path, i.e., with the angle becoming very small: $P^+ \to \eta$.
Subtraction of this self-energy part is presented graphically in
Fig.\ 1. Note that there is no need to introduce additional parameters in this
subtraction and this procedure doesn't affect the rapidity evolution equations.
Moreover, it has a clear physical interpretation: only an irrelevant contribution due to the self-energy of the light-like gauge links is removed, which is only an artifact of the unobservable background.

\begin{figure}[h]
  \includegraphics[scale=0.6,angle=90]{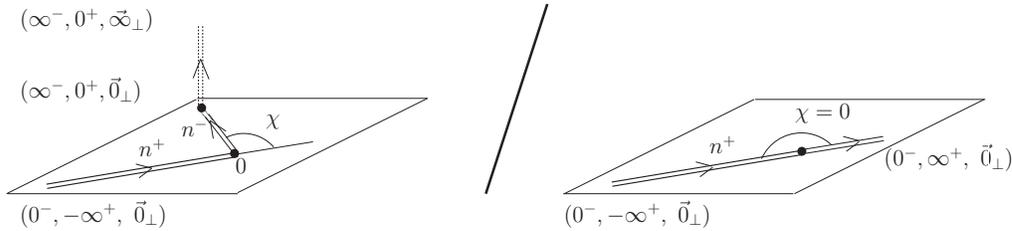}
  \caption{Subtraction of the Wilson-line self-energy contribution in
  the soft factor.}
\end{figure}

Therefore, the following generalized definition of the
${\rm A}_{\rm n}$-TMD is proposed \cite{CS_all, SC09, CKS10}
\begin{eqnarray}
  && {\cal F}_{\rm i/h}^{[{\rm A}_{\rm n}]} \left(x, {\mathbf k}_\perp; \mu, \eta\right)
=
  \frac{1}{2}
  \int \frac{d\xi^- d^2{\mathbf \xi}_\perp}{2\pi (2\pi)^2}
  {\rm e}^{-ik^{+}\xi^{-} +i {\mathbf k}_\perp
\cdot {\mathbf \xi}_\perp}
  \left\langle
              h |\bar \psi_i (\xi^-, {\mathbf \xi}_\perp)
              [\xi^-, {\mathbf \xi}_\perp;
   \infty^-, {\mathbf \xi}_\perp]^\dagger  \right.  \nonumber \\
   && \left.
\times
   [\infty^-, {\mathbf \xi}_\perp;
   \infty^-, {\mathbf \infty}_\perp]^\dagger
   \gamma^+[\infty^-, {\mathbf \infty}_\perp;
   \infty^-, \mathbf{0}_\perp]
   [\infty^-, \mathbf{0}_\perp; 0^-,\mathbf{0}_\perp]
   \psi_i (0^-,\mathbf{0}_\perp) | h
   \right\rangle
    R^{-1} \ ,
   \label{eq:general}
\end{eqnarray}
\begin{eqnarray}
&& R^{-1}(\mu, \eta)
= \nonumber \\
&& \frac{\langle 0
  |   \ {\cal P}
  \exp\Big[ig \int_{\mathcal{C}_{\rm cusp}}\! d\zeta^\mu
           \ {\cal A}^\mu (\zeta)
      \Big] \cdot
  {\cal P}^{-1}
  \exp\Big[- ig \int_{\mathcal{C'}_{\rm cusp}}\! d\zeta^\mu
           \ {\cal A}^\mu (\xi + \zeta)
      \Big]
  {| 0
  \rangle } }
{ \langle 0
  |   \ {\cal P}
  \exp\Big[ig \int_{\mathcal{C}_{\rm smooth}}\! d\zeta^\mu
           \ {\cal A}^\mu (\zeta)
      \Big] \cdot
  {\cal P}^{-1}
  \exp\Big[- ig \int_{\mathcal{C'}_{\rm smooth}}\! d\zeta^\mu
           \ {\cal A}^\mu (\xi + \zeta)
      \Big]
  | 0
  \rangle  } \ , \nonumber
\end{eqnarray}
where the gauge invariance is ensured by the inserted gauge links
$
  { [y,x] }
=
  {\cal P} \exp
  \left[-ig\int_{x}^{y}dz_{\mu} {\cal A}^\mu (z)
  \right]
$
with ${\cal A} \equiv t^a A^a$, and the contours for the soft factor are presented in Fig.\ 1.

It is worth mentioning that within the A$_{\rm n}$-TMD, the soft factor can be interpreted
in terms of the ``intrinsic'' phase resembling the Coulomb phase found by Jakob and Stefanis
in Ref. \cite{JS91} in QED.  Its origin was ascribed in \cite{JS91} to the long-range
interaction of the charged particle with its oppositely charged
counterpart that was removed ``behind the moon'' after their
primordial separation. Such a reminiscence of the quantum entanglement has been used recently \cite{CS_CMS}
in our analysis of the long-rapidity-range two-particle correlations produced at the LHC
(reported by the CMS collaboration \cite{CMS_2010_ridge}, see also these Proceedings). Therefore, within the
A$_{\rm n}$-TMD approach, one may understand the appearance of the soft factors in the TMD factorization not only in technical terms, but also from the basic principles of a quantum theory.


\paragraph{Acknowledgments}
The results reported in this work have been obtained in a long-term collaboration with N. G. Stefanis, to whom I wish to express my sincere gratitude. I am indebted to the Organizers of the ISMD 2010 for their kind hospitality and financial support. I thank the members of the theory group of Universit$\grave{\rm a}$ della Calabria (Cosenza) for their kind hospitality during my stay there with an INFN Fellowship,  and the Institute of Nuclear Theory (Seattle, WA) for the support of my visit within the program
``Gluons and the quark sea at high energies: distributions, polarization, tomography'', where this work has been launched. I thank I. Anikin, A. Bacchetta, V. Braun, M. Burkardt, F. Ceccopieri, J. Collins, L. Gamberg, A. Metz, S. Mikhailov, D. M\"uller, B. Musch, A. Prokudin, N. Stefanis, O. Teryaev, C. Weiss and F. Yuan for fruitful discussions.


\begin{footnotesize}

\end{footnotesize}

\end{document}